\documentstyle [12pt,eqsecnum,aps] {revtex}
\input epsf
\newcommand{\beq}{\begin{equation}}
\newcommand{\eeq}{\end{equation}}
\newcommand{\beqs}{\begin{eqnarray}}
\newcommand{\eeqs}{\end{eqnarray}}

\begin{document}

\draft

\baselineskip 6.0mm

\title{Large-$q$ series expansion for the ground state degeneracy of 
the $q$-state Potts antiferromagnet on the $(3 \cdot 12^2)$ lattice}

\vspace{4mm}

\author{Shan-Ho Tsai\thanks{Electronic address: tsai@insti.physics.sunysb.edu}}

\address{
Institute for Theoretical Physics  \\
State University of New York       \\
Stony Brook, N. Y. 11794-3840}

\maketitle

\vspace{4mm}

\begin{abstract}

  We calculate the large-$q$ series expansion for the 
ground state degeneracy (= exponent of the ground state entropy) per site 
of the $q$-state Potts antiferromagnet on the $(3 \cdot 12^2)$ lattice, to 
order $O(y^{19})$, where $y=1/(q-1)$. We note a remarkable agreement, to 
$O(y^{18})$, between this series and a rigorous lower bound derived recently.

\end{abstract}


\vspace{10mm}

\pagestyle{empty}
\newpage

\pagestyle{plain}
\pagenumbering{arabic}
\renewcommand{\thefootnote}{\arabic{footnote}}
\setcounter{footnote}{0}

\section{Introduction}

    Nonzero ground state disorder and associated entropy, $S_0 \ne 0$, is an
important subject in statistical mechanics.  One physical example is provided
by ice, for which the residual molar entropy is 
$S_0 = 0.82 \pm 0.05$ cal/(K-mole), i.e., $S_0/R = 0.41 \pm 0.03$, where 
$R=N_{Avog.}k_B$ \cite{ice,lp}.  A particularly simple model exhibiting 
ground state entropy without the complication of frustration is the $q$-state 
Potts antiferromagnet (AF) \cite{potts,wurev} on a lattice $\Lambda$, for 
$q \ge \chi(\Lambda)$, where $\chi(\Lambda)$ denotes the minimum number of
colors necessary to color the vertices of the lattice such that no two adjacent
vertices have the same color.  This model has a deep connection with 
graph theory in mathematics, since the zero-temperature partition function of 
the above-mentioned $q$-state Potts antiferromagnet on a lattice $\Lambda$ 
satisfies $Z(\Lambda,q,T=0)_{PAF}=P(\Lambda,q)$, where $P(G,q)$ is the 
chromatic polynomial \cite{rtrev} 
expressing the number of ways of coloring the vertices of a graph 
$G$ with $q$ colors such that no two adjacent vertices (connected by a bond of
the graph) have the same color. Hence, the ground state entropy per site is
given by $S_0/k_B = \ln W(\Lambda,q)$, where $W(\Lambda,q)$, the ground state
degeneracy per site, is 
\beq 
W(\Lambda,q) = \lim_{n \to \infty} P(\Lambda_n,q)^{1/n}
\label{w}
\eeq
Here, $\Lambda_n$ denotes an $n$-vertex lattice of type $\Lambda$ 
with appropriate (e.g., free) boundary conditions. Since nontrivial exact 
solutions for this function are known in only a very few cases (square 
lattice for $q=3$ \cite{lieb}, triangular lattice \cite{baxter}, and 
kagom\'e lattice for $q=3$ \cite{baxter70,wurev}), it is important to exploit 
and extend general approximate methods that can be applied to all cases.  
Such methods include rigorous upper and lower 
bounds, large-$q$ series expansions, and Monte Carlo measurements.  Recently,
with R. Shrock, the present author studied the ground state entropy in 
antiferromagnetic Potts models on 
various lattices and obtained further results with these three methods 
\cite{p3afhc}-\cite{warch}. We derived a general lower bound 
on $W(\Lambda,q)$ \cite{warch} which applies to all Archimedean lattices 
and coincides to many orders with large-$q$ series expansion of this function.
Previous large-$q$ series expansions include works by Baker \cite{baker}, 
Nagle \cite{nagle1,nagle}, Kim and Enting \cite{ke}, Bakaev et al 
\cite{bakaevetal} and work reported in \cite{w3,warch}. Related work on
series expansions for the ground state degeneracy of ice was done by Nagle
\cite{nagleice}.

  Large-$q$ series expansions of the respective $W(\Lambda,q)$ functions on
various Archimedean lattices were computed in Ref. \cite{warch}.  In
particular, $W(\Lambda,q)$ for $\Lambda = (3 \cdot 12^2)$ was computed to 
$O(y^{13})$ \cite{warch}.  In the present paper we extend this series
to higher order, namely to $O(y^{19})$. Our main motivation is to check the
accuracy of the lower bound on $W((3 \cdot 12^2),q)$ given in \cite{warch}.
It is interesting that this lower bound coincides with the first 19 terms, 
i.e. to $O(y^{18})$, in the large-$q$ series. We choose the lattice
$\Lambda = (3 \cdot 12^2)$ as an illustrative example of a heteropolygonal
Archimedean lattice. 
The reader is referred to Refs. \cite{p3afhc}-\cite{warch} for further 
background and references.

\section{Large-$\lowercase{q}$ series expansion}

\begin{figure}
\centering
\leavevmode
\epsfxsize=4.0in
\epsffile{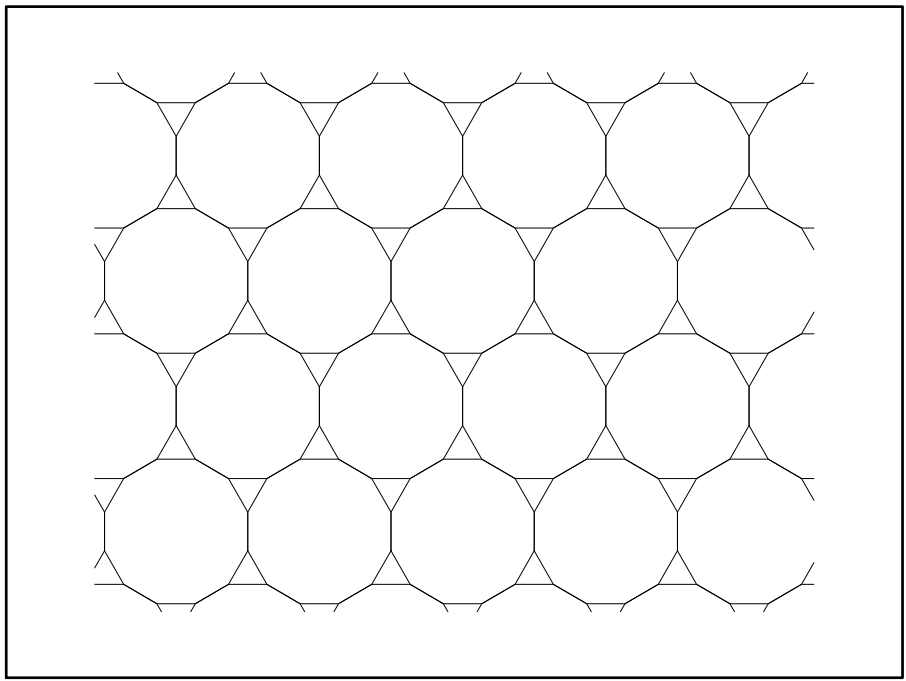}
\caption{Section of the $(3 \cdot 12^2 )$ Archimedean lattice}
\label{fig31212}
\end{figure}

    Before proceeding, we recall that an Archimedean lattice is defined as 
a uniform tiling of the plane by regular polygons in which all vertices 
are equivalent \cite{gsbook}.  Such a lattice is specified by the ordered 
sequence of polygons that one traverses in
making a complete circuit around a vertex in a given (say counterclockwise)
direction.  This is incorporated in the mathematical notation for an 
Archimedean lattice $\Lambda$: 
\beq
\Lambda = (\prod_i p_i^{a_i}) 
\label{lambda}
\eeq
where in the above circuit, the notation $p_i^{a_i}$ indicates that the 
regular polygon $p_i$ occurs contiguously $a_i$ times; it can also occur
noncontiguously. We shall denote $a_{i,s}$ as the sum of the $a_i$'s over 
all of the occurrences of the given $p_i$ in the product.  Because the 
starting point is irrelevant, the symbol is 
invariant under cyclic permutations. The number of polygons of type $p_i$ 
per site is given by
\beq
\nu_{p_i} = \frac{a_{i,s}}{p_i}
\label{nu}
\eeq
The coordination number for an Archimedean lattice is 
$\Delta = \sum_i a_{i,s}$.
In particular, for the $(3 \cdot 12^2)$ lattice considered in this paper, 
the number of triangles per site is $p_3=1/3$, the
number of $12$-gons per site is $p_{12}=1/6$, and the coordination number
is $\Delta=3$. A section of this lattice is shown in Fig.(\ref{fig31212}). 

    A general upper bound on a chromatic polynomial for an $n$-vertex
graph $G$ is $P(G,q) \le q^n$.  This yields the corresponding upper bound 
$W(\{G\},q) < q$.  Hence, as in our previous work \cite{p3afhc}-\cite{warch}, 
it is natural to define a 
reduced function that has a finite limit as $q \to \infty$, 
\beq
W_r(\{G\},q) = q^{-1}W(\{G\},q)
\label{wr}
\eeq
When calculating large-$q$ Taylor series expansions for $W$ functions on
regular lattices , it is most convenient to carry this out for the related
function 
\beq
\overline W(\Lambda,y) = \frac{W(\Lambda,q)}{q(1-q^{-1})^{\Delta/2}} 
\label{wbar}
\eeq
for which the large-$q$ series can be written in the form 
\beq
\overline W(\Lambda,y)=1+\sum_{m=1}^\infty w_{\Lambda,m} y^m
\label{wseries}
\eeq
with 
\beq
y = \frac{1}{q-1}
\label{y}
\eeq

Our calculations of large-$q$ series expansion use the method of 
Ref. \cite{nagle}. The chromatic polynomial is written as the sum
\beq
P(G,q)=\frac{(q-1)^E}{q^{(E-n)}}\sum_{G_a\leq G} (-1)^e \frac{q^{(e-v)}}
{(q-1)^e} m(G_a,q), 
\label{series}
\eeq
where $E$ is the total number of edges of the $n$-vertex graph $G$,
$m(G_a,q)$ are weights \cite{fn1} of weak subgraphs $G_a$ of $G$, and $e$ 
and $v$ are the numbers of edges and vertices, respectively, of $G_a$. 
The summation is over all weak subgraphs $G_a$. The weight function $m(G_a,q)$
vanishes if $G_a$ has any vertices of degree one or if $G_a$ has a bridge. 
Another property is that $m(G_a,q)$ does not change under the insertion or
deletion of vertices of degree two in $G_a$. Thus in the summation of 
weak subgraphs, one effectively has to consider (connected and disconnected)
subgraphs with no vertices of degree one and without bridges.
One also has to consider graphs with articulation points. Weight functions 
$m(G_a,q)$ are independent of $G$ and satisfy a simple recursion relation 
formula.  Ref. \cite{nagle} gives the weights of all star graphs with 
cyclomatic number less or equal to four.\cite{fne}

\begin{figure}
\centering
\leavevmode
\epsfxsize=4.0in
\epsffile{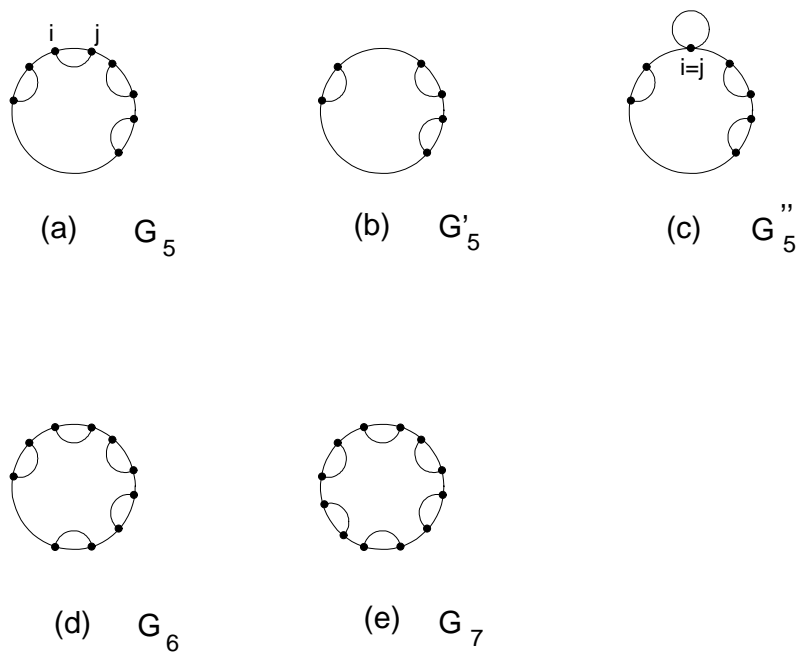}
\caption{Illustration of star graphs with cyclomatic number $c \le 7$. 
(a) $G_5$ is a graph with $c=5$; (b) $G_5'$ is derived from $G_5$ 
by omitting one edge between vertices $i$ and $j$, and (c) $G_5''$ is the 
graph with vertices $i$ and $j$ identified. (d) $G_6$ is a graph with $c=6$
and (e) $G_7$ has $c=7$. Vertices of degree two are not shown. }
\label{circle}
\end{figure}

The series expansion for $W((3 \cdot 12^2),q)$ to $O(y^{19})$ involves 
star graphs with cyclomatic number up to 7. The graphs with cyclomatic 
number equal to 5, 6 and 7 which enter in the series expansion to this order 
are shown in 
Figs.(\ref{circle}a), (\ref{circle}d) and (\ref{circle}e), respectively. 
To derive the weights of these graphs, we use Theorems III and VII of 
Ref. \cite{nagle}. Theorem III states that if a graph $G$ consists of two
pieces $G_1$ and $G_2$ which have just one vertex in common, its weight 
is given by
\beq
m(G,q)=\frac{1}{q}m(G_1,q)m(G_2,q).
\label{articul}
\eeq
Theorem VII states that 
\beq
m(G,q)=-\frac{1}{q}m(G',q)+m(G'',q),
\label{recursion}
\eeq
where $G'$ is derived from $G$ by omitting the edge between two vertices, say 
$i$ and $j$, and $G''$ is the graph with vertices $i$ and $j$ identified. 
As an example, consider the graphs $G_5$, $G_5'$ and $G_5''$ depicted in 
Figs.(\ref{circle}a), (\ref{circle}b) and (\ref{circle}c).  $G_5''$ has an
articulation point and, using Eq.(\ref{articul}), we can write its
weight as $m(G_5'',q)=q^{-1}m(P,q)m(G_5',q)$, where $P$ here stands for 
polygon. The weights for $P$ and $G_5'$ are 
$m(P,q)=(q-1)$ and $m(G_5',q)=(q-1)(q-2)^3/q^3$ \cite{nagle}.
Hence, Eq.(\ref{recursion}) yields 
\beq
m(G_5,q)=-\frac{1}{q}m(G_5',q)+m(G_5'',q)=\frac{1}{q^4}(q-1)(q-2)^4.
\eeq
Note that vertices of degree two have been omitted in Fig.(\ref{circle}).

The weights of the graphs with higher cyclomatic numbers, shown in 
Figs.(\ref{circle}d) and (\ref{circle}e), can be similarly determined to be
\beq
m(G_6,q)=\frac{1}{q^5}(q-1)(q-2)^5,
\eeq
and
\beq
m(G_7,q)=\frac{1}{q^6}(q-1)(q-2)^6,
\eeq
respectively.

\begin{figure}
\centering
\leavevmode
\epsfxsize=4.0in
\epsffile{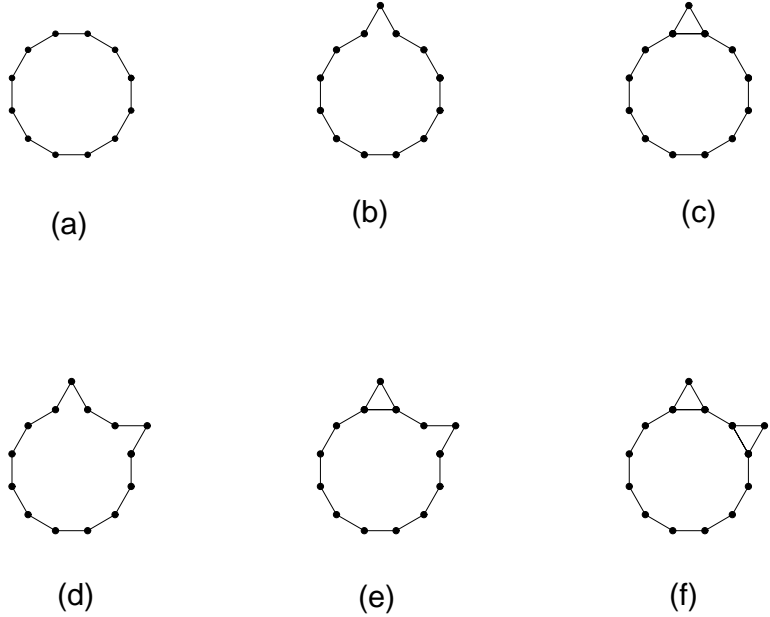}
\caption{Illustration of some graphs which enter in the large-$q$ series 
for $W((3 \cdot 12^2),q)$. Graphs (a), (b) and (c) enter in series to 
$O(y^{12})$, while graphs (d), (e) and (f) contribute
to $O(y^{13})$ and higher.}
\label{graph31212} 
\end{figure}

The subgraphs which contribute to the series to $O(y^{12})$ are (i) graphs
formed by $t$ disconnected triangles, where $t=1,2,...,6$, (ii) polygons
with $12$ vertices ($12$-gons), as shown in Fig.(\ref{graph31212}a), 
(iii) graphs formed by a triangle connected to a $12$-gon, as shown in
Figs.(\ref{graph31212}b) and (\ref{graph31212}c). Let us refer to the edge
in the overlap between a triangle and a $12$-gon as an internal edge. 
Internal edges can be part of the subgraph (as in Fig.(\ref{graph31212}c))
or not (as in Fig.(\ref{graph31212}b)). 

Further subgraphs which contribute to $O(y^{13})$ are (i) graphs formed by 
one $12$-gon and one disconnected triangle and (ii) graphs formed by
two triangles connected to a $12$-gon. In the latter case, one has to consider
graphs with
no internal edges (as in Fig.(\ref{graph31212}d)), one internal edge
(as in Fig.(\ref{graph31212}e)) and two internal edges 
(as in Fig.(\ref{graph31212}f)). Moreover, one has to consider all the
distinguishable permutations in the positions of the two triangles. In the
remaining of this paper, when we refer to $t$ triangles connected to $12$-gons,
we are including all possible distinguishable permutations of the triangles
and all cases where $i$ internal edges belong to the subgraph, with
$i=0,1,...,t$.

To $O(y^{14})$ other subgraphs which enter in the series are (i) graphs
formed by 7 disconnected triangles, (ii) graphs formed by
3 triangles connected to a $12$-gon,
(iii) graphs formed by one triangle connected to a $12$-gon and one
disconnected triangle. 

Series to $O(y^{15})$ includes (i) graphs formed by 4 triangles connected
to a $12$-gon, (ii) graphs formed by 2 triangles connected to a $12$-gon
and one disconnected triangle, (iii) two disconnected triangles and one 
$12$-gon.

To $O(y^{16})$, extra subgraphs are (i) graphs comprised of 8 disconnected
triangles, (ii) graphs formed by 5 triangles connected to a $12$-gon,
(iii) graphs composed of a triangle connected to a $12$-gon and two
disconnected triangles, (iv) graphs formed by 3 triangles connected to a
$12$-gon and one disconnected triangle.

To $O(y^{17})$, other subgraphs are (i) graphs formed by 6 triangles 
connected to a $12$-gon, (ii) 3 disconnected triangles and one $12$-gon,
(iii) 2 triangles connected to a $12$-gon and two disconnected triangles,
(iv) 4 triangles connected to a $12$-gon and one disconnected triangle.

To $O(y^{18})$, further subgraphs are (i) graphs
formed by 9 disconnected triangles, (ii) graphs formed by one triangle
connected to a $12$-gon and 3 disconnected triangles, (iii) graphs comprised
of 3 triangles connected to a $12$-gon and two disconnected triangles, (iv)
graphs formed by 5 triangles connected to a $12$-gon and one disconnected 
triangle.

To $O(y^{19})$, new subgraphs are (i) graphs formed by 4 disconnected
triangles and one $12$-gon, (ii) graphs formed by 2 triangles connected
to a $12$-gon and 3 disconnected triangles, (iii) graphs formed by 4 triangles
connected to a $12$-gon and 2 disconnected triangles, (iv) graphs formed by 6 
triangles connected to a $12$-gon and one disconnected triangle, (v) $20$-gons.

To this order, we obtain  

$$\overline W((3 \cdot 12^2),y) = 1 - \frac{1}{3}y^2 - \frac{1}{9}y^4
- \frac{5}{3^4}y^6 - \frac{10}{3^5}y^8 - \frac{22}{3^6}y^{10} +
\frac{1}{6}y^{11} - \frac{154}{3^8}y^{12} $$
\beq
- \frac{1}{18}y^{13}
 - \frac{374}{3^9}y^{14}  - \frac{1}{54}y^{15}  - \frac{935}{3^{10}}y^{16}
 - \frac{5}{486}y^{17}  - \frac{21505}{3^{13}}y^{18} + \frac{719}{1458}y^{19}
+ O(y^{20})
\label{wbar31212taylor}
\eeq

The lower bound of Ref. \cite{warch}, namely 
\beq
\overline W((3 \cdot 12^2),y)_{l}=(1-y^2)^{1/3}(1+y^{11})^{1/6},
\label{wlb}
\eeq
coincides with the first 19 terms of
the series given in Eq.(\ref{wbar31212taylor}), i.e. to $O(y^{18})$. 
This is remarkable and shows that this lower bound is indeed a 
very accurate approximation to the exact solution for the $\overline W$
function. The lower bound first differs from the large-$q$ expansion for the
exact $\overline W$ function at order $y^{19}$: the Taylor series
expansion of this lower bound gives $-(5/729)y^{19}$ whereas the
large-$q$ series expansion of $\overline W$ yields $(719/1458)y^{19}$.

It is interesting to note that the lowest order in $y$ in which the 
bound (\ref{wlb}) differs from the series is an order in which subgraphs
involving two adjacent $12$-gons, i.e. $20$-gons, first contribute in the 
series expansion. If one were to calculate the series expansion without
considering the contribution of $20$-gons to $O(y^{19})$, one would get
a result which coincides with the coefficient of the $O(y^{19})$ term
in the Taylor series of the bound (\ref{wlb}). 

\section{Conclusions}

We report on large-$q$ series expansion for the ground state degeneracy
of the Potts antiferromagnet on the $(3 \cdot 12^2)$ lattice, to $O(y^{19})$.
It is remarkable that the lower bound derived previously coincides with
the first 19 terms of the series, i.e. to $O(y^{18})$.  
It is worthwhile to perform similar series expansions to high orders
for other types of lattices. 

I would like to thank Professor Robert Shrock for helpful comments on the 
manuscript.

This research was supported in part by the NSF grant PHY-97-9722101.

\vfill
\eject

\begin{thebibliography}{99}
\bibitem{fn1}{In Ref.\cite{nagle} the weights $m(G_a,q)$ are called 
reduced weights and denoted by $w(G_a,q)$. A total weight is also defined
therein. However, in the present paper we shall not refer to the total
weight.}
\end{thebibliography}
\end{document}